%% file: ms.tex
\documentclass[a4paper,fleqn,usenatbib]{mnras}
\usepackage{graphicx}
\usepackage{amssymb}
\usepackage{amsmath}
\usepackage[T1]{fontenc}
\usepackage{ae,aecompl}
\usepackage{natbib,times}
\usepackage{lscape}
\usepackage{array}
\usepackage{setspace}
\usepackage{helvet}
\usepackage{float}
\newcommand{\PreserveBackslash}[1]{\let\temp=\\#1\let\\=\temp}
\newcolumntype{C}[1]{>{\PreserveBackslash\centering}p{#1}}
\newcolumntype{R}[1]{>{\PreserveBackslash\raggedleft}p{#1}}
\newcolumntype{L}[1]{>{\PreserveBackslash\raggedright}p{#1}}

\begin{document}

\title[The BCG alignment] {The alignment between brightest cluster
  galaxies and host clusters}

\author[Yuan \& Wen]
       {Z. S. Yuan,$^{1,2}$ \thanks{E-mail: zsyuan@nao.cas.cn}
        and Z. L. Wen$^{1,2}$ 
\\
1. National Astronomical Observatories, Chinese Academy of Sciences, 
20A Datun Road, Chaoyang District, Beijing 100101, China\\
2. CAS Key Laboratory of FAST, NAOC, Chinese Academy of Sciences,
           Beijing 100101, China}

\date{Accepted XXX. Received YYY; in original form ZZZ}

\label{firstpage}
\pagerange{\pageref{firstpage}--\pageref{lastpage}}
\maketitle


\begin{abstract}
The alignment between brightest cluster galaxies (BCGs) and host
clusters can reveal the mystery of formation and evolution for galaxy
clusters. We measure cluster orientations in optical based on the
projected distribution of member galaxies and in X-ray by fitting the
morphology of intra-cluster medium (ICM). Cluster orientations
determined in the two wavelengths are generally consistent. The
orientation alignment between BCGs and host clusters is confirmed and
more significant than previous works. We find that BCGs are more
aligned with cluster orientations measured in X-ray than those from
optical data. Clusters with a brighter BCG generally show a stronger
alignment. We argue that the detected redshift evolution of the
alignment is probably caused by observational bias rather than
intrinsic evolution. The alignment is not related to the ellipticity
of BCGs, and the richness, ellipticity and dynamical state of host
clusters. The strong alignment between BCGs and morphology of ICMs may
be the consequence of the co-evolution between the central massive
galaxy and host clusters.
\end{abstract}


\section{Introduction}
\label{intro}
Galaxy clusters are the largest gravitationally bound systems in the
Universe, formed by accreting materials along filaments and merging
smaller groups or subclusters
\citep[e.g.,][]{ps74,b91,swj+05,mbb+09}. The anisotropic processes for
mass assembly lead clusters and their member galaxies orient
non-randomly \citep[e.g.,][]{d98,acc06,fjl+08,kyk+08}. Satellite
galaxies in groups or clusters tend to point to the central massive
galaxies \citep[e.g.,][]{t76,pk05,yvm+06,flm+07,hmf+18}. BCGs are
observed to align with their host clusters \citep[hereafter ``BCG
  alignment'' or ``the alignment'' for short,
  e.g.,][]{s68,nsd+10,hkf+11,hmf+16,wdbp17,wfg19}. On larger scales,
galaxy groups or clusters also orient toward their neighbours
statistically \citep[e.g.,][]{b82,rk87,w89,wpy+09,smbn12}.
Investigations on the orientation alignment of galaxies, groups and
clusters provide a unique way to understand their formation and
evolution histories \citep[see reviews:][]{jck+15,kcj+15,kbh+15}.

BCGs are the most massive galaxies in clusters, generally inhabit the
dense environment of cluster centers. They gain masses through
accreting materials from the ICM or cannibalizing satellite galaxies
\citep[e.g.,][]{oh75,f94,d98,ngd18}, and show peculiar properties in
many aspects when compared to normal elliptical galaxies
\citep[e.g.,][]{bwh81,bvk+07,lom+10,sym+14,wh15a}. On the one hand,
the properties of BCGs are affected by the ambient environment
\citep[e.g.,][]{mhrc09,kvc+15,yhw16}; on the other hand, BCGs can
regulate the distribution of surrounding gases and heat them
significantly through AGN feedback \citep[e.g.,][]{cfb+09,mmn13}. The
orientation alignment indicates the tight links between BCGs and host
clusters.

The formation of the BCG alignment is still controversial: whether the
alignment is created primordially and maintained to now, or it appears
gradually during the process of mass assembly of BCGs. This could be
partly answered by comparing the alignment signal in different
redshifts. \citet{nsd+10} and \citet{hkf+11} found stronger alignment
for lower redshift clusters, but \citet{hmf+16} claimed that no clear
redshift evolution is detected. By using the {\it Hubble Space
  Telescope} ({\it HST}), \citet{wdbp17} found that the BCG alignment
has been clearly set ten billion years ago.

Generally, the mass distribution of galaxy clusters can be well traced
by the distribution of their member galaxies
\citep[e.g.,][]{bbc+05,zbu+09,zbb+11}. The overall orientation of
clusters thus can be measured with the positions of member galaxies
\citep[e.g.,][]{s68,cm80}. Clear BCG alignment has been detected for
large cluster samples \citep[e.g.,][]{nsd+10,hkf+11,hmf+16}. Relations
between the alignment strength and cluster properties are also
discussed. For example, clusters with a more luminous or more dominant
BCG usually show more remarkable alignments
\citep[e.g.,][]{nsd+10,hkf+11,hmf+16}. While marginal or no dependence
is found on the richness of clusters \citep[e.g.,][]{hkf+11,hmf+16}.

The cluster mass can also be reflected by the distribution of ICM
\citep[e.g.,][]{frp84,asf02}, thus the global orientation of galaxy
clusters can be determined from their X-ray morphologies. Based on
{\it Einstein} images, \citet{psh91} found clear BCG alignments with
41 clusters, and \citet{rl91} got consistent results for 26
clusters. \citet{afe+95} detected the BCG alignment in 5 clusters
observed by the {\it ROSAT}. \citet{hhb08} studied the alignment for
30 clusters with {\it Chandra} images, in much higher angular
resolution than the {\it Einstein} and {\it ROSAT}
images. \citet{der+16} confirmed the BCG alignment for 25 clusters, no
matter the cluster orientation is measured from X-ray ({\it Chandra}),
Sunyaev-Zel'dovich effect ({\it Bolocam}) or lensing ({\it HST})
data. All the above X-ray samples are very limited, insufficient to
further investigate the influence of cluster properties on the BCG
alignment.

The {\it Chandra} and {\it XMM-Newton} satellites have accumulated
massive data for galaxy clusters. Recently, we processed the archival
{\it Chandra} \citep{yh20} and {\it XMM-Newton} \citep{yhw22} data
homogeneously, and obtained X-ray images for 1844 clusters in
total. It is feasible now to measure the orientation of clusters in
both optical and X-ray wavelengths for a large sample of clusters. In
Section 2, we describe the cluster samples and derive the shape
parameters. Results on the BCG alignment are presented in Section 3. A
summary is given in Section 4. Throughout this paper, we assume a flat
$\Lambda$CDM cosmology with $H_0=70$ km~s$^{-1}$ Mpc$^{-1}$,
$\Omega_m=0.3$ and $\Omega_{\Lambda}=0.7$.

\begin{figure}
\centering
\includegraphics[width=0.3\textwidth, angle=0]{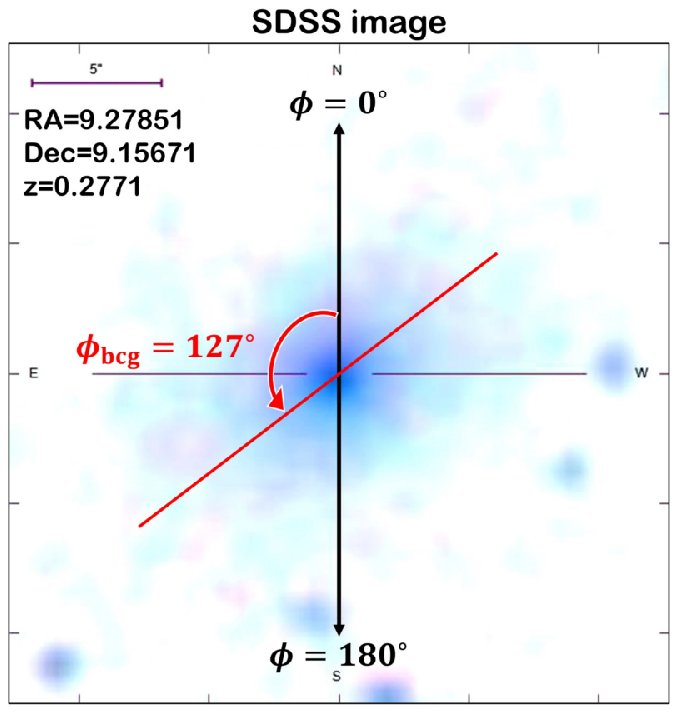}\\[3mm]
\includegraphics[width=0.31\textwidth, angle=-90]{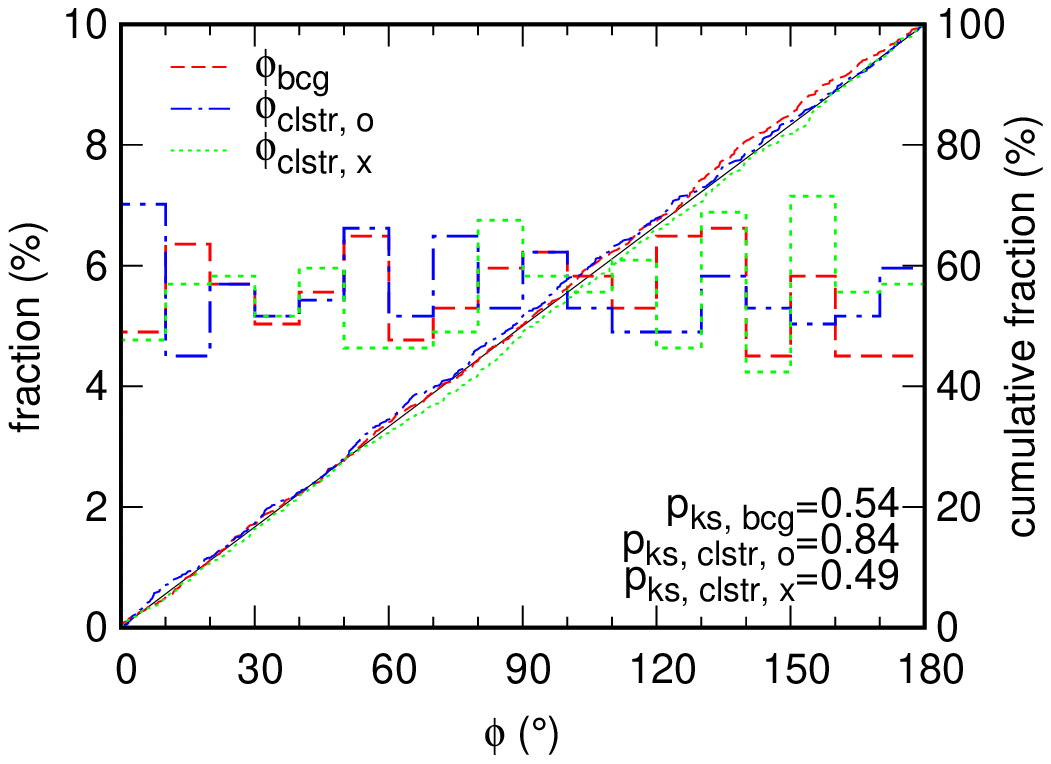}
\caption{Upper panel: the observational image for the BCG of Abell 68
  obtained by the {\it SDSS} (contrast enhanced). The coordinate and
  redshift of the BCG are written on the top-left corner. The modeled
  position angle $\phi_{\rm bcg}$ is labeled, from the north through
  the east. Lower panel: distributions of $\phi_{\rm bcg}$ (dashed),
  $\phi_{\rm clstr,~o}$ (dash-dotted) and $\phi_{\rm clstr,~x}$
  (dotted) for the 755 overlapped clusters. Histograms ($x$ and $y1$
  axes) indicate the fraction of position angles in each bin, oblique
  curves ($x$ and $y2$ axes) mean the cumulative fractions. The
  straight-solid line is the theoretical line for random
  distribution. The probability of KS-test between the curve of
  $\phi_{\rm bcg}$ ($\phi_{\rm clstr,~o}$ or $\phi_{\rm clstr,~x}$)
  and the solid line is written in the bottom-right corner.}
\label{fig1}
\end{figure}

\section{Data}
\label{data}
\subsection{Cluster samples}

The cluster sample in X-ray is derived from the archival data of the
{\it Chandra} and {\it XMM-Newton} \citep{yh20,yhw22}. Galaxy clusters
are collected in two independent approaches: 1) clusters from targeted
observations, and 2) those serendipitously detected. The X-ray images
of clusters are processed homogeneously, and smoothed to a certain
physical scale of 30 kpc \citep[refer][for details on the sample
  collection and image procssing]{yh20,yhw22}. Finally, 964 clusters
are obtained from {\it Chandra} images \citep{yh20}, 1308 clusters are
from {\it XMM-Newton} images, and a joint sample with 1844 clusters is
derived by combining the {\it Chandra} and {\it XMM-Newton} samples
\citep{yhw22}.

In optical, the large cluster catalogue derived by \citet{wh15b} is
used. \citet{whl12} identified 132\,684 clusters from the photometric
data of {\it Sloan Digital Sky Survey} ({\it SDSS}). \citet{wh15b}
updated this catalogue based on the {\it SDSS} spectroscopic data
\citep{aaa+15}, and identified 25\,419 new clusters. In total, the
sample in \citet{wh15b} contains 158\,103 clusters.

We cross-match the X-ray joint sample in \citet{yhw22} with the
optical catalogue in \citet{wh15b} within the cluster radius
$r_{500}$, the radius of a cluster within which the mean matter
density is 500 times of the local critical density. Preliminarily, we
get 755 clusters overlapped in the X-ray and optical samples.

\subsection{Parameters of galaxy clusters}
\subsubsection{Orientation and ellipticity of BCGs}
Based on the BCG coordinates in the catalogue of \citet{wh15b}, we
directly get the position angle $\phi_{\rm bcg}$ ({\tt deVPhi\_r}) and
the axis ratio $b/a$ ({\tt deVAB\_r}) of BCGs, fitted with the de
Vancouleurs model, from the {\it SDSS} database
\citep{slb+02}.\footnote{http://skyserver.sdss.org/dr15/en/tools/crossid/crossid.aspx}
The BCG ellipticity $\varepsilon_{\rm bcg}$ is defined as
\begin{equation}
\varepsilon_{\rm bcg}=\frac{1-b/a}{1+b/a}.
\label{phibcg}
\end{equation}
The $\phi_{\rm bcg}$ is measured from the north through the east
within $0^\circ\le\phi_{\rm bcg}<180^\circ$. As an example, the upper
panel of Fig.~\ref{fig1} shows the {\it SDSS} image for the BCG of
Abell 68 and the fitted position angle. In the lower panel of
Fig.~\ref{fig1}, we present the distribution of $\phi_{\rm bcg}$
(dashed) for the 755 overlapped clusters, and find it follows the
random distribution quite well.

\begin{figure}
\centering
\includegraphics[width=0.35\textwidth, angle=0]{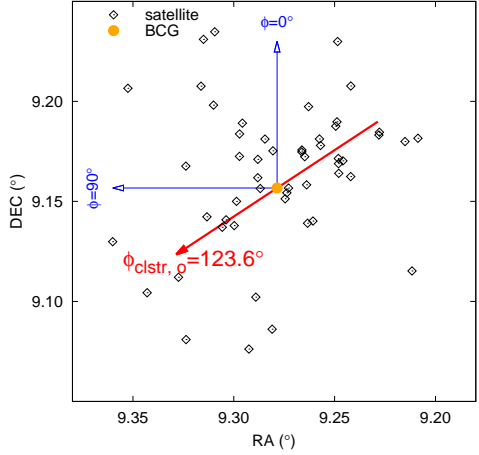}
\caption{The distribution of member galaxies for the Abell 68 and the
  derived position angle $\phi_{\rm clstr,~o}$. The member galaxies
  are taken from \citet{wh15b} and selected with $M_{\rm r}\le-20.5$
  and $r<r_{500}$.}
\label{fig2}
\end{figure}

\subsubsection{Orientation and ellipticity of clusters estimated from member galaxies}
The orientation of clusters can be derived from the distribution of
member galaxies \citep[e.g.,][]{kas+02,nsd+10,hmf+16}. First, we
calculate the three moments:
\begin{equation}
\begin{split}
& M_{xx}=\frac{1}{N_{\rm sat}}\sum\limits_{i}\frac{x_{i}^2}{r_{i}^2},\\
& M_{xy}=\frac{1}{N_{\rm sat}}\sum\limits_{i}\frac{x_{i}y_{i}}{r_{i}^2},\\
& M_{yy}=\frac{1}{N_{\rm sat}}\sum\limits_{i}\frac{y_{i}^2}{r_{i}^2},\\
\end{split}
\end{equation}
where $x_{i}$, $y_{i}$ are the relative angular distances in the right
ascension and declination directions, respectively, of the $i$th
satellite galaxy to the BCG, $r_{i}$ is equal to
$\sqrt{x_{i}^2+y_{i}^2}$, and $N_{\rm sat}$ is the number of satellite
galaxies in the cluster. Here, we only adopt satellite galaxies
satisfying: (1) the $r$-band absolute magnitude is brighter than
$-20.5$ mag, and (2) the projected distance to the BCG is less than
$r_{500}$. The two parameters $Q$ and $U$ are defined as
\begin{equation}
\begin{split}
& Q=M_{xx}-M_{yy},\\
& U=2M_{xy}. 
\end{split}
\end{equation}
Then the position angle and the ellipticity of clusters can be
computed through
\begin{equation}
\phi_{\rm clstr,~o}=\frac{1}{2}{\rm arctan}\left(\frac{U}{Q}\right), 
\end{equation}
\begin{equation}
\varepsilon_{\rm clstr,~o}=\sqrt{Q^2+U^2}.
\end{equation}
The $\phi_{\rm clstr,~o}$ is measured from the north through the east
within $0^\circ\le\phi_{\rm clstr,~o}<180^\circ$. In Fig.~\ref{fig2},
the distribution of member galaxies for the Abell 68 and the
corresponding position angle $\phi_{\rm clstr, o}$ are presented as an
example. In the lower panel of Fig.~\ref{fig1}, we present the
distribution of $\phi_{\rm clstr, o}$ for 755 clusters and find that
the cluster orientations (dash-dotted) are randomly distributed in the
projected plane.

\begin{figure}
\centering
\includegraphics[width=0.35\textwidth, angle=0]{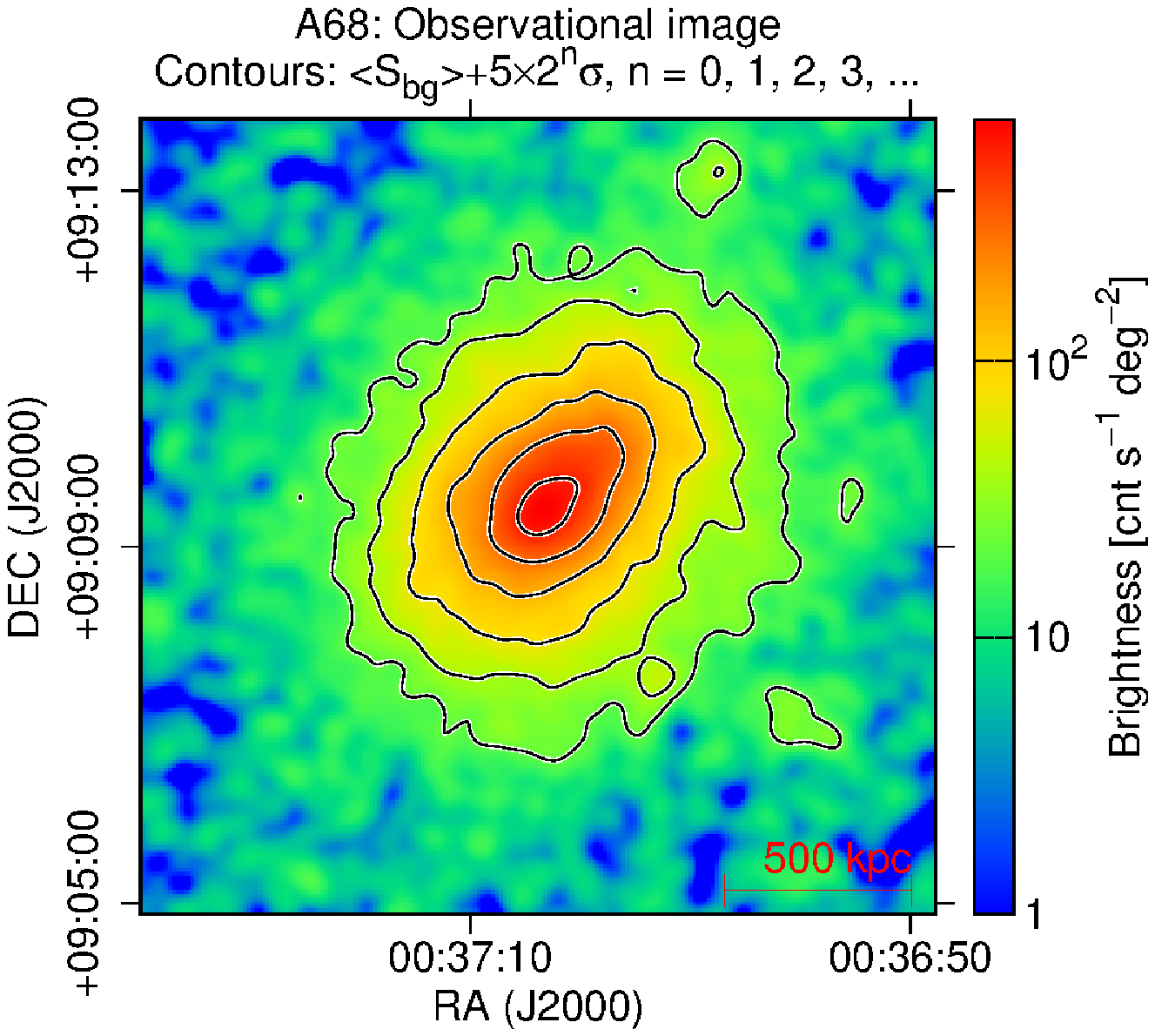}\\[3mm]
\includegraphics[width=0.35\textwidth, angle=0]{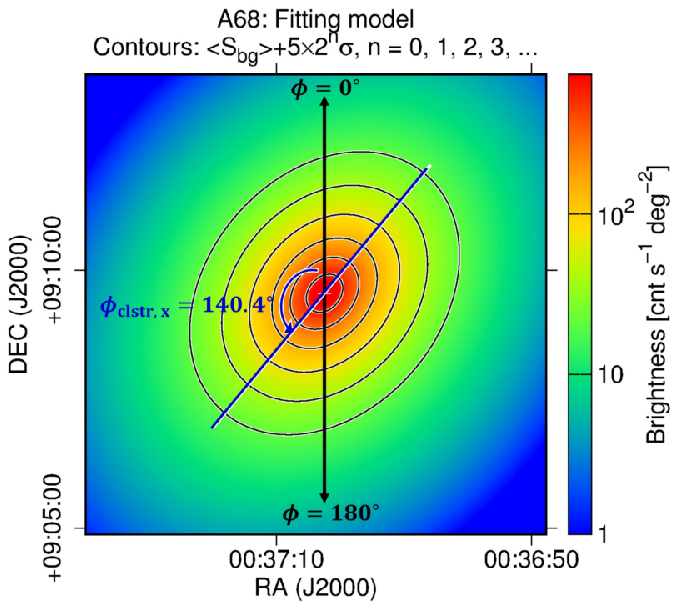}
\caption{Upper panel: the observational X-ray image of Abell 68
  obtained by the {\it XMM-Newton}. The scale of 500 kpc is labeled
  on the bottom-right corner. Lower panel: the best-fitted
  $\beta$-model for the X-ray image with the position angle $\phi_{\rm
    clstr,~x}$ marked.}
\label{fig3}
\end{figure}

The deviation angle between the BCG and host cluster thus can be
calculated through
\begin{equation}
\begin{cases}
\Phi_{\rm o}=|\phi_{\rm bcg}-\phi_{\rm clstr,~o}|~~~~~~~~~~~~~~~~(|\phi_{\rm
  bcg}-\phi_{\rm clstr,~o}|\le90^{\circ}),\\ 
\Phi_{\rm o}=180^{\circ}-|\phi_{\rm bcg}-\phi_{\rm clstr,~o}|~~~(|\phi_{\rm
  bcg}-\phi_{\rm clstr,~o}|>90^{\circ}).
\end{cases}
\label{Phio}
\end{equation}
Here the $\Phi_{\rm o}$ is in $0^\circ\le\Phi_{\rm o}\le90^\circ$. 

\begin{figure*}
\centering
\includegraphics[width=0.3\textwidth, angle=-90]{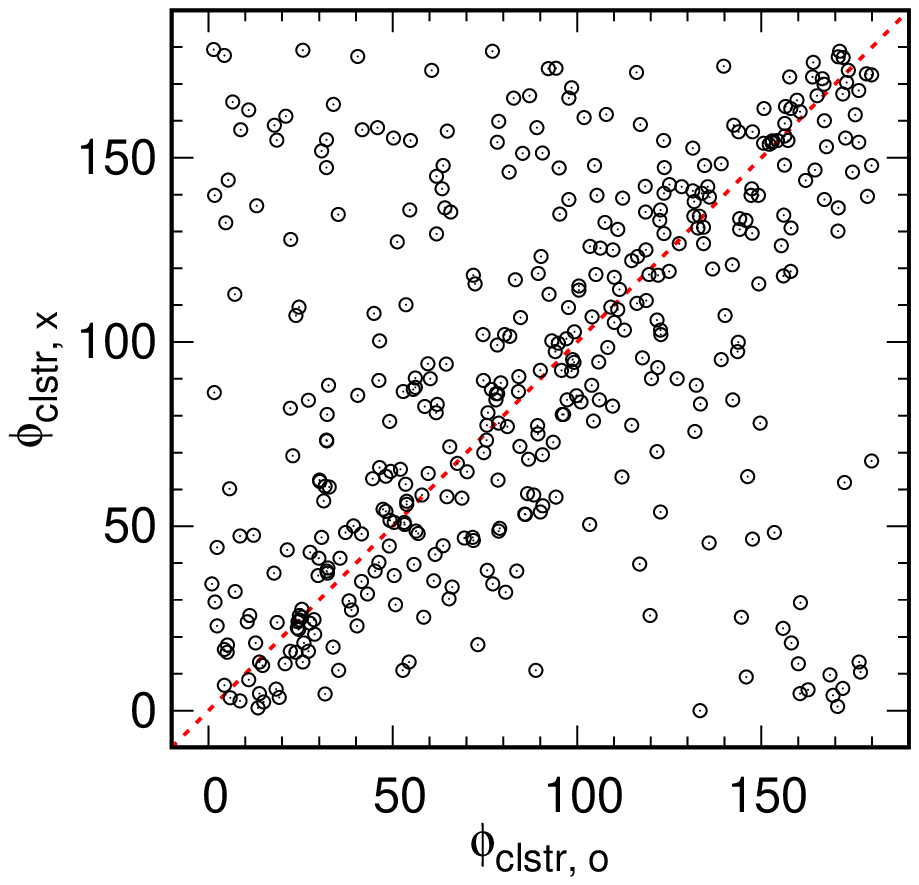}
\hspace{2.2cm}
\includegraphics[width=0.3\textwidth, angle=-90]{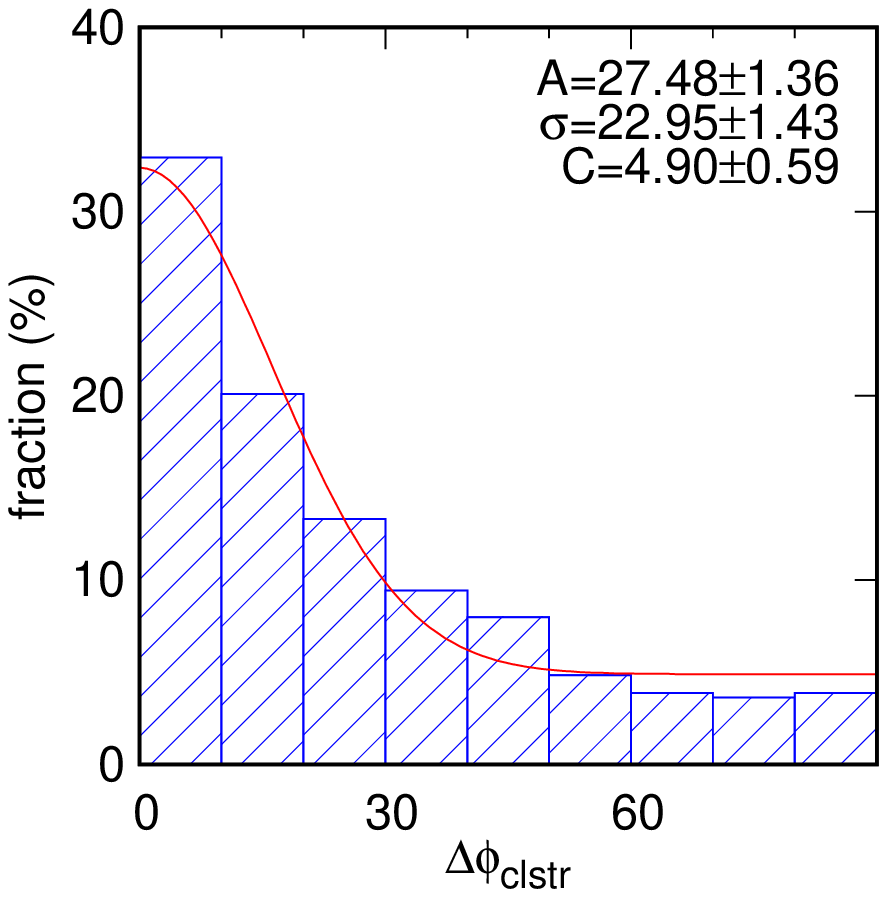}\\[3mm]
\includegraphics[width=0.3\textwidth, angle=-90]{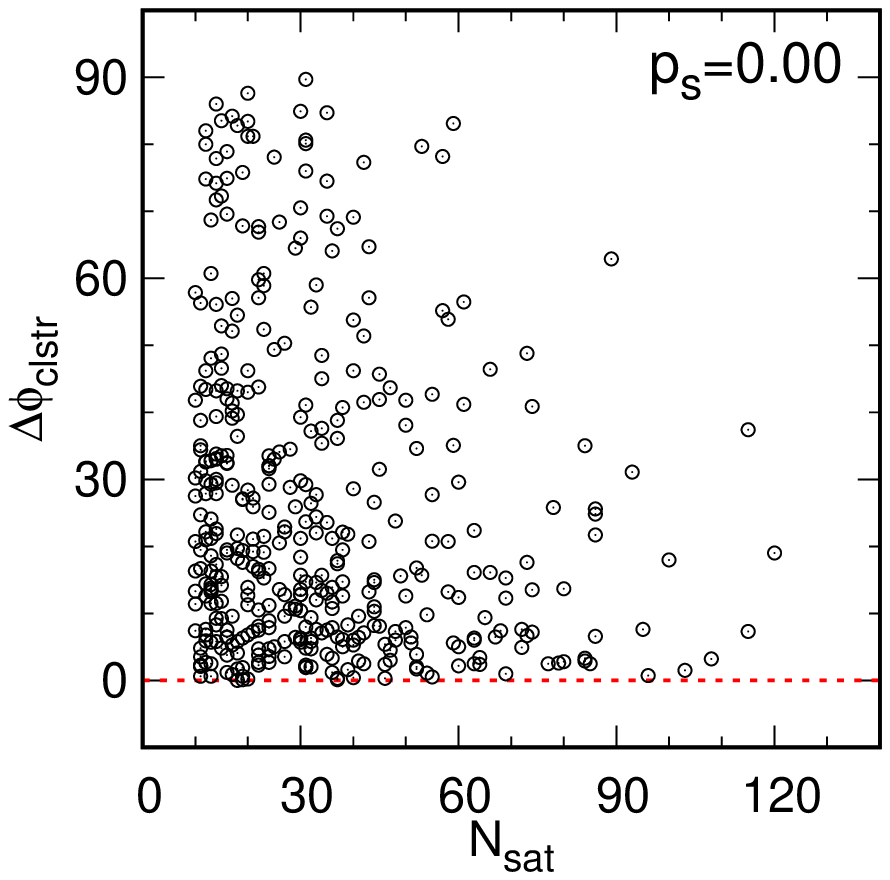}
\hspace{2cm}
\includegraphics[width=0.3\textwidth, angle=-90]{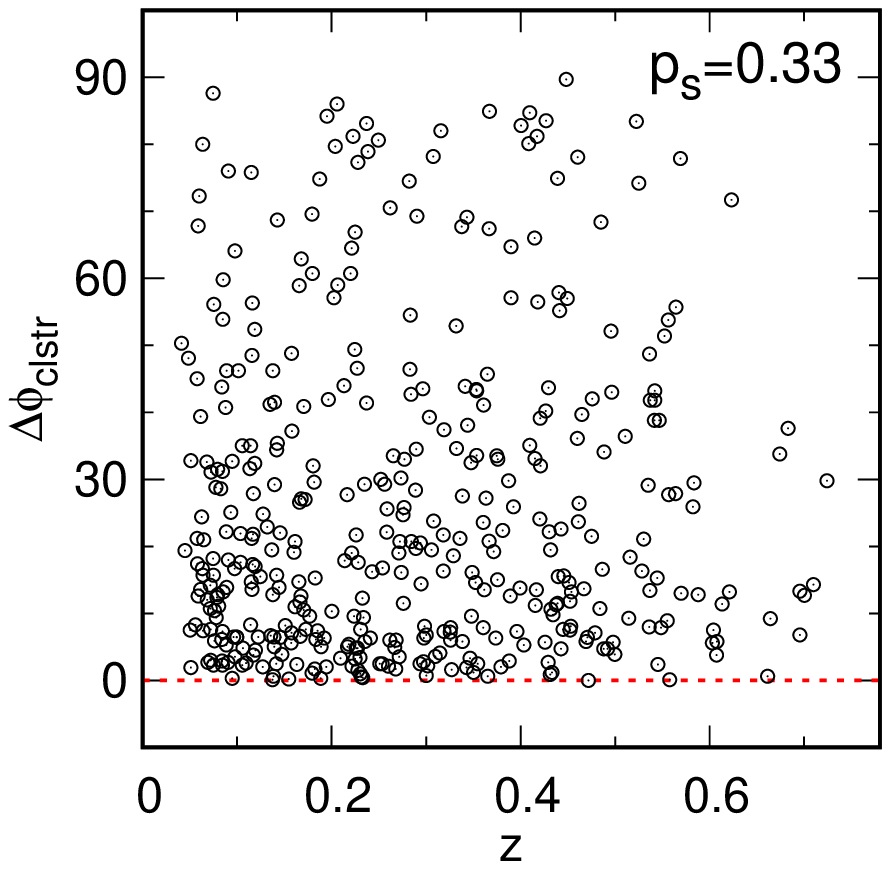}
\caption{Upper-left panel: comparison of cluster orientations
  estimated from optical ($\phi_{\rm clstr,~o}$) and X-ray ($\phi_{\rm
    clstr,~x}$) data. The dotted line indicates the
  equivalence. Upper-right panel: the distribution for differences
  between $\phi_{\rm clstr,~o}$ and $\phi_{\rm clstr,~x}$. The solid
  curve is the fitted Gaussian model for the distribution. Parameters
  of the fitted model are written on the top-right corner. Lower-left
  panel: the relation between the position angle difference
  $\Delta\phi_{\rm clstr}$ and the number of satellite galaxies in
  clusters $N_{\rm sat}$. The dotted line means $\Delta\phi_{\rm
    clstr}=0$. The value of the significance of Spearman rank-order
  correlation $p_{\rm s}$ is labeled on the top-right
  corner. Lower-right panel: similar to the bottom-left panel but for
  the relation between $\Delta\phi_{\rm clstr}$ and redshift $z$.}
\label{fig4}
\end{figure*}

\subsubsection{Orientation and ellipticity of clusters estimated from X-ray images}
The orientation and ellipticity of host clusters can also be measured
by fitting the cluster X-ray images with a two-dimensional
$\beta$-model as \citep[see] [for details]{yh20}
\begin{equation}
f(x,y)=f(r)=\mathbb{A}(1+(r/r_0)^2)^{-\beta}+\mathbb{C},
\label{fx}
\end{equation}
where
\begin{equation}
r(x,y)=\frac{\sqrt{x'^2(1-\varepsilon_{\rm clstr,~x})^2+y'^2}}{1-\varepsilon_{\rm clstr,~x}},
\label{r}
\end{equation}
and
\begin{equation}
\begin{split}
x'=(x-x_0)\cos\phi_{\rm clstr,~x}+(y-y_0)\sin\phi_{\rm clstr,~x},\\
y'=(y-y_0)\cos\phi_{\rm clstr,~x}-(x-x_0)\sin\phi_{\rm clstr,~x}.
\label{xy}
\end{split}
\end{equation}
Here $\varepsilon_{\rm clstr,~x}$ is the cluster ellipticity,
$\phi_{\rm clstr,~x}$ is the position angle from the north through the
east within $0^\circ\le\phi_{\rm clstr,~x}<180^\circ$. The upper panel
of Fig.~\ref{fig3} shows the X-ray image of Abell 68 observed by the
{\it XMM-Newton}, refer \citet{yhw22} for details. In the lower panel,
the corresponding $\beta$-model and the fitted cluster orientation
$\phi_{\rm clstr,~x}$ are presented. The distribution of $\phi_{\rm
  clstr,~x}$ (dotted) for 755 clusters is shown in the lower panel of
Fig.~\ref{fig1}, follows the random distribution perfectly.

The deviation angle between orientations of the BCG and host cluster
thus can be calculated with
\begin{equation}
\begin{cases}
\Phi_{\rm x}=|\phi_{\rm bcg}-\phi_{\rm clstr,~x}|~~~~~~~~~~~~~~~~(|\phi_{\rm
  bcg}-\phi_{\rm clstr,~x}|\le90^{\circ}),\\ 
\Phi_{\rm x}=180^{\circ}-|\phi_{\rm bcg}-\phi_{\rm clstr,~x}|~~~(|\phi_{\rm
  bcg}-\phi_{\rm clstr,~x}|>90^{\circ}),
\end{cases}
\label{Phix}
\end{equation}
where the $\Phi_{\rm x}$ is in the range of $0^\circ\le\Phi_{\rm
  x}\le90^\circ$.

Since the uncertainty of measured orientations is larger for rounder
sources \citep[e.g., see][]{rka+12}, previous works usually set a
threshold for the ellipticity \citep[e.g.,][]{yvm+06,nsd+10}. In this
paper, we limit ellipticities, including the $\varepsilon_{\rm bcg}$,
$\varepsilon_{\rm clstr,~o}$ and $\varepsilon_{\rm clstr,~x}$, larger
than $0.1$. Meanwhile, because the measurement of the orientation
$\phi_{\rm clstr,~o}$ is inaccurate for clusters with few members, we
only use $\phi_{\rm clstr,~o}$ that estimated from clusters with 10 or
more satellites, i.e., $N_{\rm sat}\ge10$. Finally, we obtain 411
clusters satisfying the above selection criteria.

\subsubsection{Other parameters}
To study the relation between BCG alignment and properties of BCGs or
host clusters, we get the BCG redshift $z$, the $r$-band absolute
magnitude of BCGs $M_{\rm bcg}$, the number of satellite galaxies
$N_{\rm sat}$, and the cluster richness $R_{\rm clstr}$ from the
catalogue in \citet{wh15b}. About 90\% of the BCG redshifts are
estimated from spectroscopic data. The absolute magnitude $M_{\rm
  bcg}$ is evolution-corrected with $M_{\rm bcg}=M_{\rm bcg}^{\rm
  obs}+1.16z$. The cluster richness $R_{\rm clstr}$ is estimated by
the total luminosity of member galaxies and can be taken as a good
mass proxy of clusters \citep{wh15b}.

In \citet{yhw22}, we calculated four kinds of dynamical parameters
with X-ray images, i.e., the concentration index $c$, the centroid
shift $\omega$, the power ratio $P_3/P_0$, and the morphology index
$\delta$. Since these parameters are tightly correlated with each
other \citep{yh20}, we only take the adaptive morphology index
$\delta$ as the dynamical parameter for clusters in following
discussions. The morphology index $\delta$ is defined as the best
combination of the profile parameter $\kappa$ and the asymmetry
factor $\alpha$ estimated from the maps of surface brightness. It is
calibrated by using a complete sample of clusters with known relaxed
and disturbed state. Clusters with $\delta<0$ are relaxed, while
those with $\delta>0$ are regarded as disturbed clusters. All the
parameters for the 411 clusters are listed in Table~\ref{tab1}.

\input{Tab1.tex}

\section{Results}
\subsection{Comparing cluster orientations estimated from optical and X-ray data}

Under the assumption of hydrostatic equilibrium in the gravitational
potential well of clusters, the distributions of member galaxies and
ICM are both regulated by the mass distribution of galaxy
clusters. Thus, it is natural to expect a general agreement between
cluster orientations determined from member galaxies and those from
the ICM. In the upper-left panel of Fig~\ref{fig4}, a good consistency
is presented between cluster orientations estimated from optical
($\phi_{\rm clstr,~o}$) and X-ray ($\phi_{\rm clstr,~x}$) data, though
few clusters present large deviations. In the upper-right panel of
Fig~\ref{fig4}, we show the distribution of the differences between
cluster orientations measured in optical and X-ray wavelengths,
$\Delta\phi_{\rm clstr}$, and fit the distribution with a Gaussian
model as
\begin{equation}
{\rm fraction} = A\times
{\rm exp}\left(-\left(\frac{\Phi-\Phi_0}{\sigma}\right)^{2}\right)+C.
\label{gauss}
\end{equation}
Here $A$ is the model amplitude, $\Phi_0$ is the model center and set
to 0, $\sigma$ is the model width, and $C$ is a constant. The typical
scatter of the $\phi_{\rm clstr,~o}-\phi_{\rm clstr,~x}$ relation is
indicated by the $\sigma$ of the Gaussian model, which is equal to
$22.95^{\circ}\pm1.43^{\circ}$. The lower-left panel of Fig~\ref{fig4}
shows the correlation between the $\Delta\phi_{\rm clstr}$ and the
number of satellite galaxies in clusters $N_{\rm sat}$. We assess the
degree of $N_{\rm sat}-\Delta\phi_{\rm clstr}$ correlation with the
significance of Spearman rank-order correlation $p_{\rm s}$
\citep[defined in][p. 640]{ptv+92}, which is a very sensitive
indicator for weak but intrinsic tendency, with the zero value for a
significant dependence while non-zero values for independence. We find
that clusters host more satellite galaxies tend to show better
agreement between the $\phi_{\rm clstr,~o}$ and $\phi_{\rm
  clstr,~x}$. In the lower-right panel of Fig~\ref{fig4}, we find that
the $\Delta\phi_{\rm clstr}$ is independent of the cluster redshift
$z$.

\subsection{The BCG alignment}

\begin{figure}
\centering
\includegraphics[width=0.34\textwidth, angle=-90]{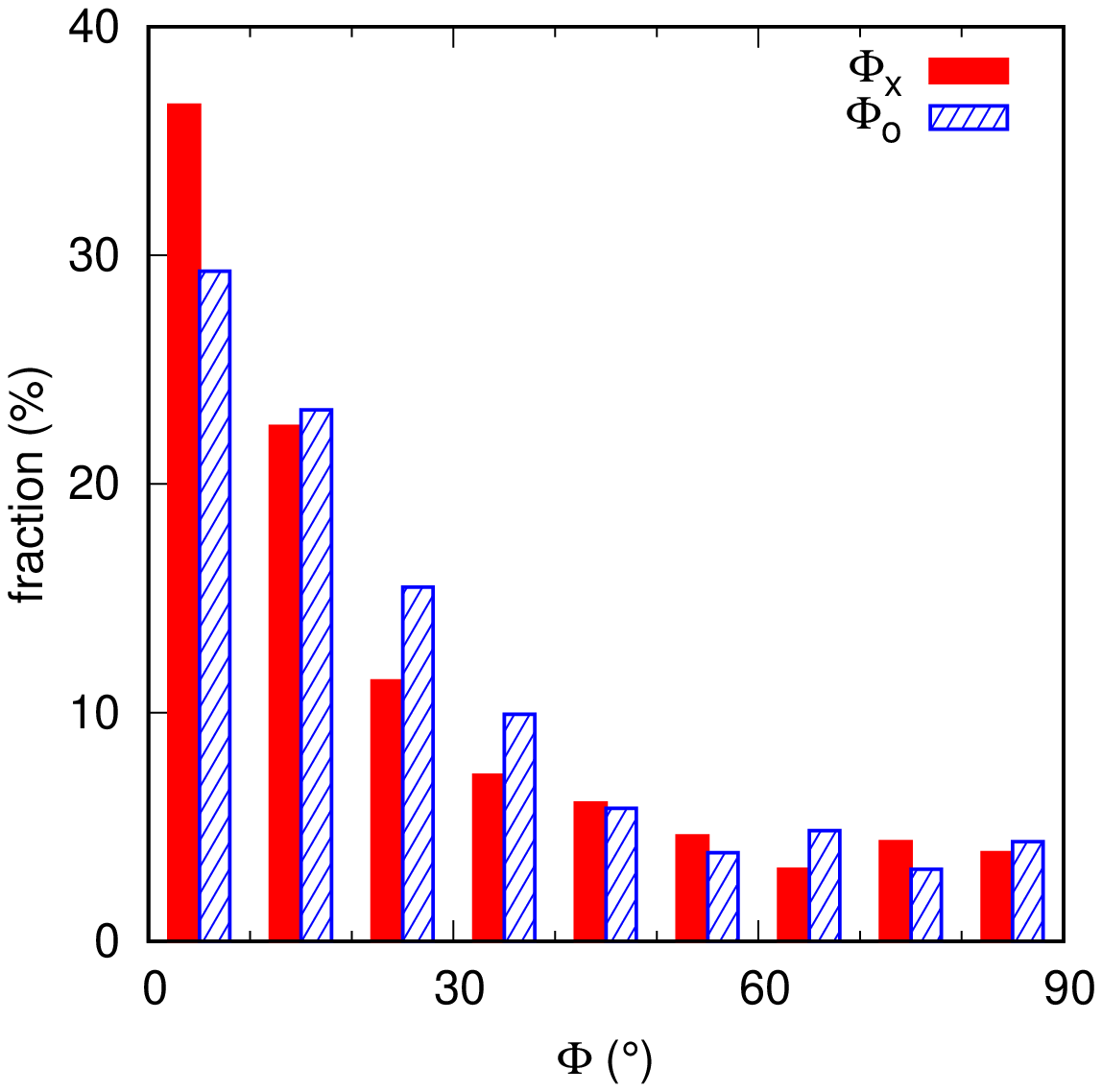}\\
\includegraphics[width=0.24\textwidth, angle=-90]{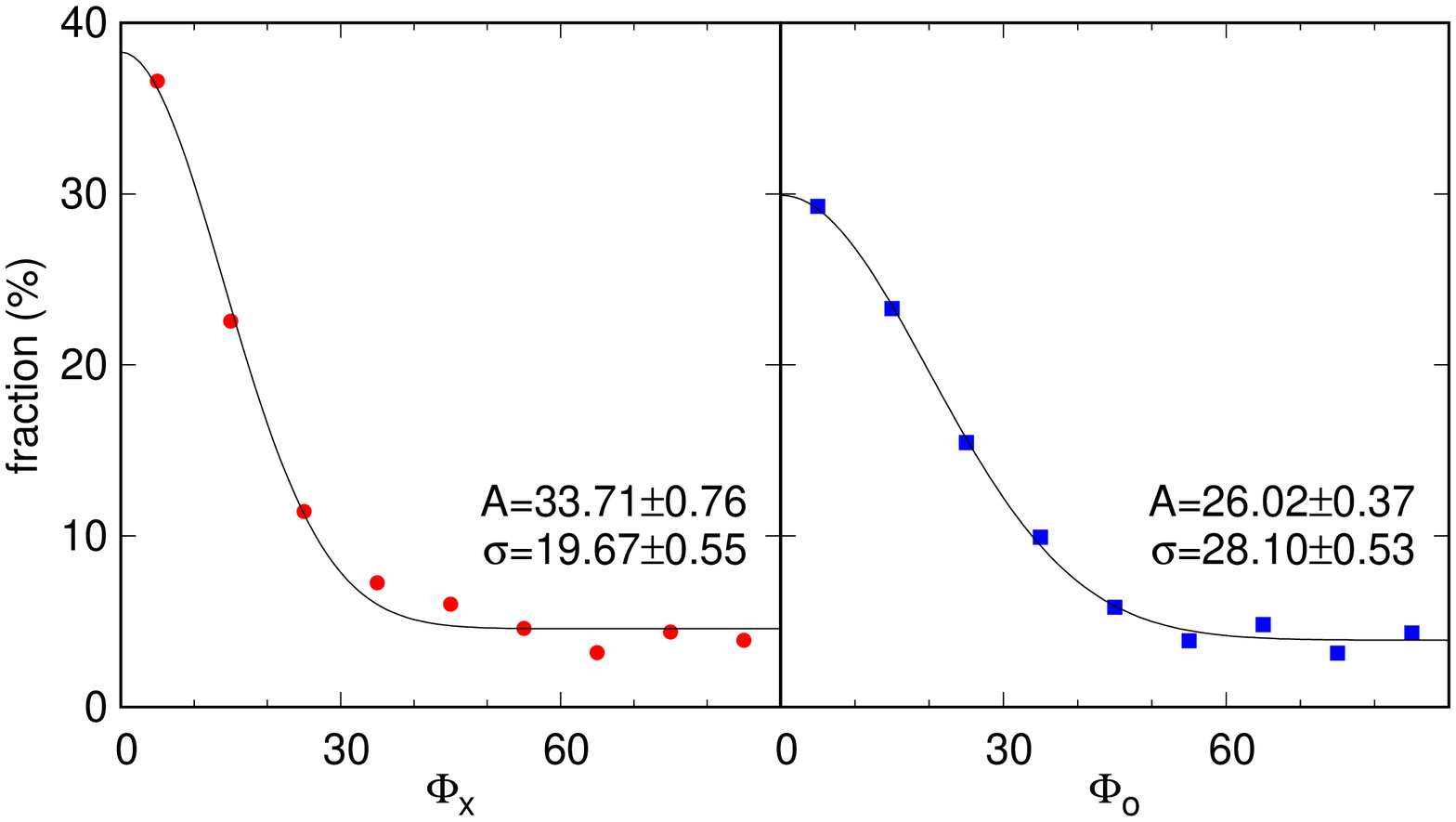}
\caption{Upper panel: distributions of the deviation angles between
  orientations of BCGs and host clusters. Lower panels: modeling the
  distributions of deviation angles with a Gaussian model. Values of
  fitted parameters are written on each panel.}
\label{fig5}
\end{figure}

\input {Tab2.tex}

\begin{figure*}
\centering
\includegraphics[width=0.75\textwidth, angle=0]{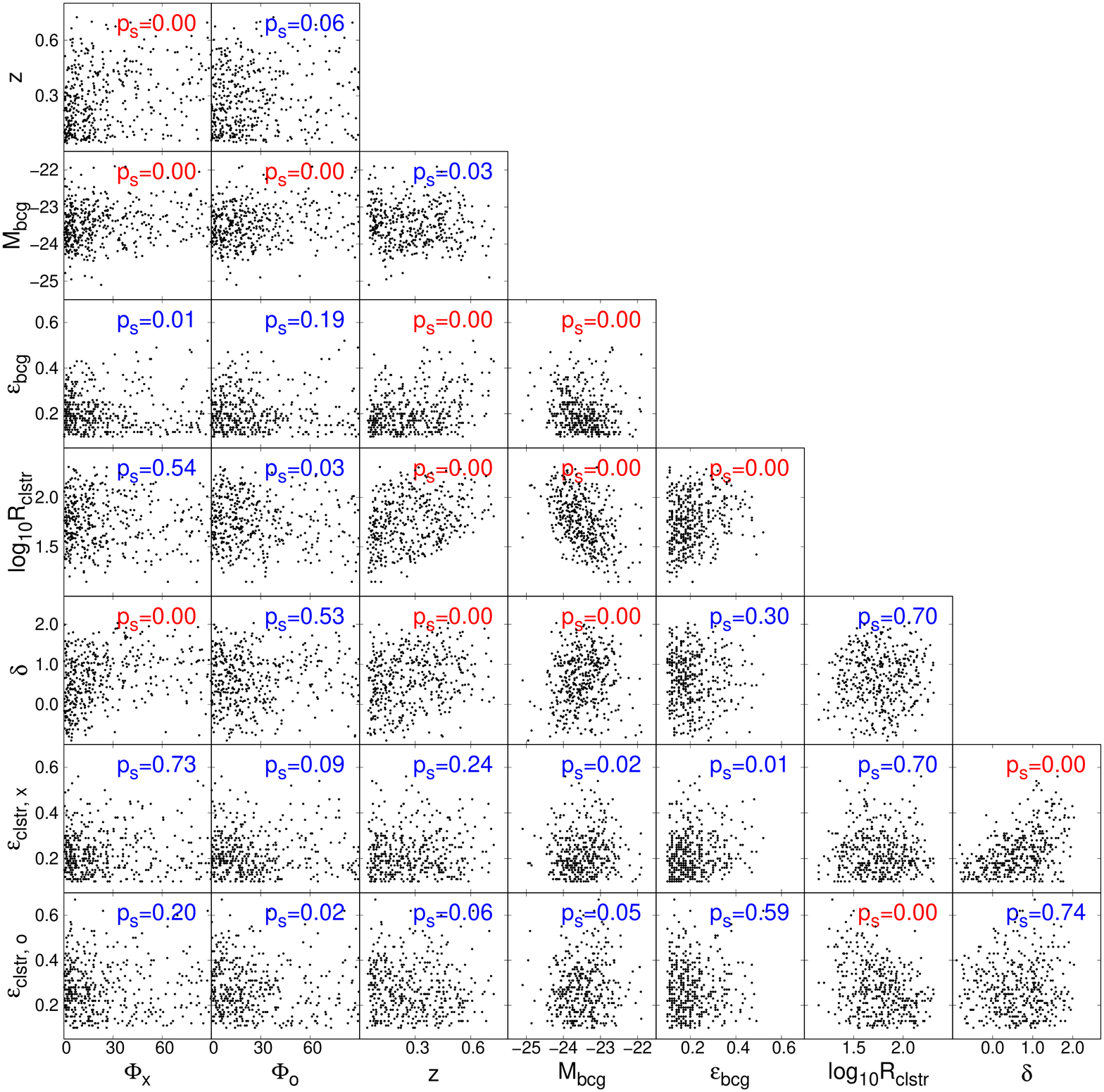}
\caption{Parameter correlations for 411 clusters. The significance of
  Spearman rank-order correlation $p_{\rm s}$ is labeled in each
  panel, the zero value in red indicates a significant correlation
  between the two parameters, while a non-zero value in blue means no
  clear correlation is found.}
\label{fig6}
\end{figure*}

In the upper panel of Fig.~\ref{fig5}, we show the distributions of
the deviation angles $\Phi$ between orientations of BCGs and host
clusters. A clear signal of alignment is shown --- the fraction of
clusters with $\Phi<30^\circ$ is significantly larger than those with
a larger $\Phi$. We find that the $\Phi_{\rm x}$ (solid) shows
stronger alignment than the $\Phi_{\rm o}$ (shaded). The distributions
of $\Phi_{\rm x}$ and $\Phi_{\rm o}$ are fitted with a Gaussian model,
as stated in Equation~\ref{gauss} and presented in the lower panels of
Fig.~\ref{fig5}. The best-fitted parameters for each samples are
listed in Table~\ref{tab2}. It is clear that the fitted model for
$\Phi_{\rm x}$ has a significantly larger amplitude $A$ and a smaller
width $\sigma$ than those of the $\Phi_{\rm o}$. This suggests that
the BCG morphology is more aligned with the distribution of ICM than
that of satellite galaxies.

In Table~\ref{tab2}, the mean deviation angles and fitting parameters
for the results obtained by \citet{nsd+10} and \citet{hmf+16} are also
presented. The alignments we obtained are significantly stronger,
i.e., larger $A$ and smaller $\sigma$ and $\langle\Phi\rangle$. This
is probably because our sample is derived from the common sample in
X-ray and optical and under stricter selection criteria, e.g., $N_{\rm
  sat}\ge10$, which leads our sample has a smaller uncertainty than
that of previous works.

\subsection{Correlations to properties of BCGs or clusters}

Correlations between the alignment strength and various properties of
BCGs or clusters have been studied, see Table~\ref{tab3} as a summary.
However, many inconsistencies are obtained by different authors. For
example, \citet{nsd+10} and \citet{hkf+11} found that the BCG
alignment is more prominent at lower redshifts, while \citet{hmf+16}
did not find this trend. \citet{hhb08} found that clusters with a more
elongated BCG generally show stronger BCG alignment, but
\citet{hmf+16} argued that the dependence on the BCG ellipticity is
probably due to selection effects.

\input{Tab3.tex}

\begin{figure*}
\centering
\includegraphics[width=0.3\textwidth, angle=-90]{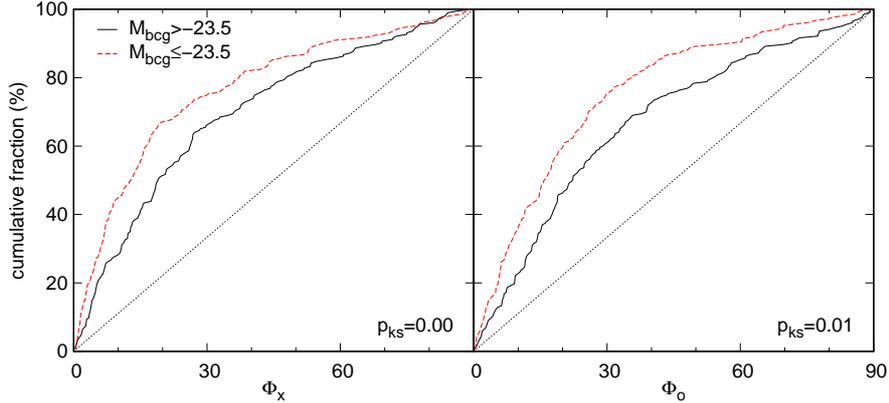}
\caption{Left panel: the cumulative distribution of the deviation
  angle $\Phi_{\rm x}$ for subsamples with different ranges of BCG
  luminosity. The probability of KS-test $p_{\rm ks}$ between the two
  subsamples is written on the bottom-right corner. The dotted line is
  the theoretical line for non-alignment. Right panel: similar to the
  left panel but for $\Phi_{\rm o}$.}
\label{fig7}
\end{figure*}

Here we search the relation between the BCG alignment and the redshift
$z$, the BCG absolute magnitude $M_{\rm bcg}$ and ellipticity
$\varepsilon_{\rm bcg}$, and the richness $R_{\rm clstr}$ and
dynamical parameter $\delta$ of host clusters. As shown in the
leftmost two columns of Fig.~\ref{fig6}, only the BCG absolute
magnitude $M_{\rm bcg}$ shows significant dependence, i.e., $p_{\rm
  s}=0$, for $\Phi_{\rm x}$ and $\Phi_{\rm o}$ simultaneously. In
Fig.~\ref{fig7}, we divide the 411 clusters into two subsamples with
the boundary of $M_{\rm bcg}=-23.5$~mag, and compare the cumulative
fraction of the two subsamples for $\Phi_{\rm x}$ (left panel) and
$\Phi_{\rm o}$ (right panel), respectively. It is clear that the
subsample with more luminous BCGs (dashed) presents stronger alignment
signal than the fainter subsample (solid). The corresponding
probability of the Kolmogorov-Smirnov (KS) test, $p_{\rm ks}$, between
the two subsamples is equal 0.00 for the $\Phi_{\rm x}$ and equal to
0.01 for the $\Phi_{\rm o}$, indicating a significantly different
distribution between the two subsamples.

\begin{figure}
\centering
\includegraphics[width=0.3\textwidth, angle=-90]{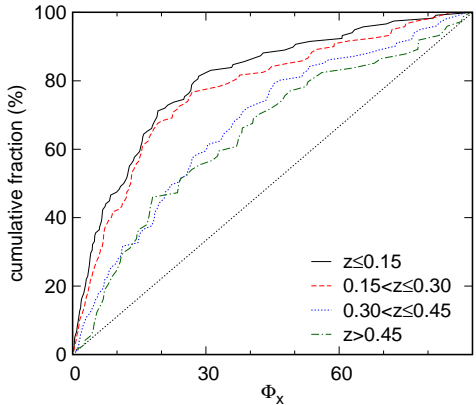}
\caption{The cumulative distribution of the deviation angle $\Phi_{\rm
    x}$ for clusters with different ranges of redshift. The dotted
  straight line is the theoretical line for non-alignment.}
\label{fig8}
\end{figure}

The redshift evolution of the BCG alignment has been discussed
\citep[e.g.,][]{nsd+10,hkf+11,hmf+16}. In Fig.~\ref{fig6}, we find a
clear dependence of $\Phi_{\rm x}$ on the cluster redshift $z$, though
the dependence vanishes for the $\Phi_{\rm o}$. Fig.~\ref{fig8} shows
the cumulative fraction of $\Phi_{\rm x}$ for subsamples with
different ranges of redshift. We find that the distributions are
similar between subsamples for $z\le0.15$ (solid) and $0.15<z\le0.30$
(dashed), but it changes significantly for the subsample with
$0.30<z\le0.45$ (dotted), and then keeps stable for $z>0.45$
(dash-dotted). Considering the discontinuous changes among subsamples
with different redshifts, we argue that the changes are not originated
from intrinsic evolution, but probably a consequence of observational
selection effect, i.e., the uncertainty of BCG orientations measured
from the {\it SDSS} images becomes larger at $z>0.3$.

The Fig.~\ref{fig6} shows the significance $p_{\rm s}$ of the
$\Phi_{\rm x}-\delta$ correlation is equal to 0, but it changes to
0.53 for the $\Phi_{\rm o}-\delta$ correlation. Considering the
inconsistency between the $\Phi_{\rm x}$ and $\Phi_{\rm o}$, we
conclude that no intrinsic relation is found between the BCG alignment
and the dynamical state of host clusters. For other parameters, i.e.,
the BCG ellipticity $\varepsilon_{\rm bcg}$, the cluster richness
$R_{\rm clstr}$ and ellipticity $\varepsilon_{\rm clstr, x}$ and
$\varepsilon_{\rm clstr, o}$, the Fig.~\ref{fig6} presents no
dependence to both $\Phi_{\rm x}$ and $\Phi_{\rm o}$. The results from
literature and that obtained by us are summarized in Table~\ref{tab3}.

\section{Summary}
Investigations on the BCG alignment and the correlations with
properties of BCGs and clusters are helpful to understand the
co-evolution of clusters and their central massive galaxies. In this
work, we combine the large X-ray sample derived by \citet{yhw22} and
the huge optical catalogue from \citet{wh15b}. The position angle of
BCGs $\phi_{\rm bcg}$ is directly taken from the {\it SDSS}
database. The orientations of host clusters are calculated in two
independent approaches: (1) the $\phi_{\rm clstr,~o}$ is computed with
the distribution of member galaxies, and (2) the $\phi_{\rm clstr,~x}$
is obtained by fitting the X-ray image of clusters with an elliptical
$\beta$-model. By setting thresholds on ellipticities and number of
satellite galaxies, finally a sample of 411 clusters are obtained.

We find that cluster orientations measured from X-ray images are
generally consistent with that determined from optical data. We
confirm the BCG alignment and find BCGs are more aligned with the
distribution of ICM compared to that of member galaxies. The alignment
signal obtained by us is much stronger than that from literature. We find
that clusters with more luminous BCGs show stronger BCG alignment
statistically. We argue that the detected changes of the BCG alignment
related to the redshift is mainly caused by selection effects rather
than intrinsic physical evolution. The BCG alignment is independent of
the BCG ellipticity, and the richness, ellipticty and dynamical state
of host clusters.

\section*{Acknowledgements}

We thank the referee, Dr. Michael West, for instructive comments
which improved the paper. We thank J. L. Han for constructive
suggestions and also Hu Zou for helpful discussions. YZS are
supported by the science research grants from the China Manned Space
Project (No. CMS-CSST-2021-A01, CMS-CSST-2021-B01). WZL acknowledges
the support by the National Natural Science Foundation of China
(12073036).
This research has made use of data obtained from the {\it Chandra Data
  Archive} and software provided by the {\it Chandra X-ray Center
  (CXC)} in the application packages {\it CIAO}, {\it ChIPS}, and {\it
  Sherpa}.
This work is based on observations obtained with {\it XMM-Newton}, an
ESA science mission with instruments and contributions directly funded
by ESA Member States and NASA. 
Funding for the Sloan Digital Sky Survey IV has been provided by the
Alfred P. Sloan Foundation, the U.S. Department of Energy Office of
Science, and the Participating Institutions. SDSS acknowledges support
and resources from the Center for High-Performance Computing at the
University of Utah. The SDSS web site is www.sdss.org.

\section*{Data availability}

The data underlying this article, including the full version of
Table~\ref{tab1} and the X-ray images for all clusters, are available
at the webpage: http://zmtt.bao.ac.cn/galaxy\_clusters/dyXimages/.

\bibliographystyle{mnras}
\bibliography{ref}
\label{lastpage}
\end{document}

%% file: Tab1.tex
\begin{table*}
\caption{Parameters for 411 clusters (the full table can be found on the webpage: http://zmtt.bao.ac.cn/galaxy\_clusters/dyXimages/).}
\renewcommand
\tabcolsep{2pt}
\footnotesize
\begin{center}          
\begin{tabular}{lrrcccrccccccccc}
\hline
  \multicolumn{1}{c}{Name} &\multicolumn{1}{c}{RA}  &\multicolumn{1}{c}{DEC} & $z$   &  \multicolumn{1}{c}{$M_{\rm bcg}$} & $\varepsilon_{\rm bcg}$ & \multicolumn{1}{c}{$\phi_{\rm bcg}$} & \multicolumn{1}{c}{$N_{\rm sat}$} &  \multicolumn{1}{c}{$R_{\rm clstr}$} &\multicolumn{1}{c}{$\delta$} & $\varepsilon_{\rm clstr,~o}$ & \multicolumn{1}{c}{$\phi_{\rm clstr,~o}$} & $\varepsilon_{\rm clstr,~x}$ & \multicolumn{1}{c}{$\phi_{\rm clstr,~x}$} & \multicolumn{1}{c}{$\Phi_{\rm o}$}  & \multicolumn{1}{c}{$\Phi_{\rm x}$}\\
 & \multicolumn{1}{c}{(J2000)}& \multicolumn{1}{c}{(J2000)} & & & &\multicolumn{1}{c}{($^\circ$)} & & & & &\multicolumn{1}{c}{($^\circ$)} & &\multicolumn{1}{c}{($^\circ$)} &\multicolumn{1}{c}{($^\circ$)}  &\multicolumn{1}{c}{($^\circ$)}\\
  \multicolumn{1}{c}{(1)} &\multicolumn{1}{c}{(2)}  &\multicolumn{1}{c}{(3)} & \multicolumn{1}{c}{(4)}   &  \multicolumn{1}{c}{(5)} & \multicolumn{1}{c}{(6)} & \multicolumn{1}{c}{(7)} & \multicolumn{1}{c}{(8)} & \multicolumn{1}{c}{(9)} & \multicolumn{1}{c}{(10)} & \multicolumn{1}{c}{(11)}  & \multicolumn{1}{c}{(12)}  & \multicolumn{1}{c}{(13)} &\multicolumn{1}{c}{(14)}  & \multicolumn{1}{c}{(15)}  & \multicolumn{1}{c}{(16)}\\
  \hline                    
A2697                  &0.79826 & -6.09169 &0.2335 &-23.70 &0.14 &163.2  &65 &105.76 &-0.27$\pm$0.01 &0.13 &147.6 &0.21 &157.0  &15.6  & 6.2\\
A2700                  &0.95698 &  2.06647 &0.0976 &-23.68 &0.24 &158.7  &36 & 71.82 &-0.33$\pm$0.01 &0.13 & 41.7 &0.26 &157.6  &63.0  & 1.1\\
WHLJ000524+161309      &1.34984 & 16.21926 &0.1156 &-23.40 &0.18 & 25.8  &34 & 63.76 & 0.76$\pm$0.01 &0.13 & 80.6 &0.30 & 32.1  &54.8  & 6.3\\
Z15                    &1.58453 & 10.86429 &0.1663 &-24.19 &0.11 &112.5  &36 & 63.25 &-0.15$\pm$0.01 &0.26 & 97.7 &0.11 &109.4  &14.8  & 3.1\\
ACTCLJ0008.1+0201      &2.04332 &  2.02009 &0.3665 &-23.87 &0.32 & 56.3  &37 & 72.60 & 1.07$\pm$0.01 &0.44 & 61.9 &0.20 &129.3  & 5.6  &73.0\\
ACTLJ0014-0056         &3.72544 & -0.95236 &0.5368 &-24.31 &0.24 & 58.3  &34 &101.23 & 0.35$\pm$0.01 &0.16 & 52.1 &0.31 & 65.5  & 6.2  & 7.2\\
CL0016+1626            &4.63993 & 16.43779 &0.5551 &-23.66 &0.42 & 61.6  &24 & 80.06 & 0.48$\pm$0.01 &0.17 & 56.9 &0.20 & 48.0  & 4.7  &13.6\\
CL0019.6+0336          &4.91107 &  3.59926 &0.2668 &-23.99 &0.25 &147.9  &72 &140.53 & 0.49$\pm$0.01 &0.35 &166.5 &0.15 &171.4  &18.6  &23.5\\
PSZ2G114.79-33.71      &5.15480 & 28.65949 &0.0949 &-23.55 &0.19 &153.8  &46 & 76.91 & 0.29$\pm$0.01 &0.29 &154.4 &0.29 &154.7  & 0.6  & 0.9\\
WHYJ003410-021039      &8.56122 & -2.08459 &0.0793 &-23.33 &0.23 & 19.9  &27 & 39.00 & 1.24$\pm$0.01 &0.29 & 13.5 &0.10 &  0.7  & 6.4  &19.2\\
\hline                  
\end{tabular}
\label{tab1}      
\end{center}
    {Columns: (1) cluster name; (2-3) right ascension and declination
      in J2000; (4) the BCG redshift; (5 - 7) the $r$-band absolute
      magnitude, ellipticity and position angle of the BCG; (8) number
      of selected satellite galaxies in the cluster; (9 - 10) the
      richness and dynamical paramter --- the morphology index of the
      cluster; (11 - 12) the ellipticity and position angle of the
      cluster measured from member galaxies; (13 - 14) the cluster
      ellipticity and position angle determined from the X-ray image; (15)
      the acute angle between $\phi_{\rm bcg}$ and $\phi_{\rm
        clstr,~o}$; (16) the acute angle between $\phi_{\rm bcg}$ and
      $\phi_{\rm clstr,~x}$.}
\end{table*}

%% file: Tab2.tex
\begin{table}
\caption{The mean departure angle and fitting parameters.}
\renewcommand\tabcolsep{3.0pt}
\begin{center}  
\begin{tabular}{l l c c c c}
\hline
  \multicolumn{1}{c}{Sample}  & \multicolumn{1}{c}{$N_{\rm sat}$} & \multicolumn{1}{c}{$\langle\Phi\rangle$}  & $A$  & $\sigma$  & $C$\\
  \hline                    
this work, $\Phi_{\rm x}$    &$\ge$10 &24.34 &33.71$\pm$0.76 &19.67$\pm$0.55 &4.58$\pm$0.29\\
this work, $\Phi_{\rm o}$    &$\ge$10 &26.15 &26.02$\pm$0.37 &28.10$\pm$0.53 &3.91$\pm$0.20\\
N10                        &$>$4  &$\sim$38$^*$ &7.86$\pm$0.20 &41.12$\pm$1.54 &7.93$\pm$0.17\\
H16                        &---  &35.07 &11.41$\pm$0.27 &33.42$\pm$1.08 &6.26$\pm$0.17\\
\hline                  
\end{tabular}
\label{tab2}      
\end{center}
    {N10 = \citet{nsd+10}, H16 = \citet{hmf+16}.\\ $^*$: estimated by
      us from the solid curve in the figure 5 of \citet{nsd+10}.}
\end{table}

%% file: Tab3.tex
\begin{table*}
\caption{A summary on the relation between BCG alignment and BCG or
  cluster properties.}  \renewcommand\tabcolsep{3.0pt}
\begin{center}  
\begin{tabular}{l c c c c c c}
\hline
  \multicolumn{1}{c}{Reference}  &\multicolumn{1}{c}{Cluster number} &  \multicolumn{1}{c}{$z$-evolution} & \multicolumn{1}{c}{BCG luminosity} & \multicolumn{1}{c}{BCG ellipticity} & \multicolumn{1}{c}{cluster richness} & \multicolumn{1}{c}{cluster dynamic} \\
  \multicolumn{1}{c}{(1)} &\multicolumn{1}{c}{(2)}  &\multicolumn{1}{c}{(3)} & \multicolumn{1}{c}{(4)}   &  \multicolumn{1}{c}{(5)} & \multicolumn{1}{c}{(6)} & \multicolumn{1}{c}{(7)} \\
  \hline                    
\citet{hhb08}           &30          &--  &--  &Yes &--       &No  \\
\citet{nsd+10}          &7031        &Yes &--  &--  &Marginal &--  \\
\citet{hkf+11}          &$\sim$11000 &Yes &Yes &--  &No       &--  \\
\citet{hmf+16}          &8237        &No  &Yes &No  &Marginal &--  \\
this work               &411         &No  &Yes &No  &No       &No  \\
\hline                  
\end{tabular}
\label{tab3}      
\end{center}
    {``--'' means the parameter is not discussed.}
\end{table*}